\documentstyle[12pt]{article}
\def\be {\begin{equation}}
\def\ee {\end{equation}}
\def\ba {\begin{eqnarray}}
\def\ea {\end{eqnarray}}

\def\bi {\begin{itemize}}
\def\ei {\end{itemize}}
\begin{document}
\def\bea{\begin{eqnarray}}
\def\eea{\end{eqnarray}}
\title{\bf Casimir stress on parallel plates in de Sitter space with  signature
change}
\author{M.R. Setare  \footnote{E-mail: rezakord@ipm.ir}
  \\{Department of Science,  Payame Noor University. Bijar. Iran}}
\date{\small{}}

\maketitle
\begin{abstract}
The Casimir stress on two parallel plates in a de Sitter background
corresponding to different metric signatures and cosmological
constants is calculated for massless scalar fields satisfying Robin
boundary conditions on the plates.  Our calculation
 shows that for the parallel plates with false vacuum between and true
  vacuum outside, the total Casimir pressure leads to an attraction of the
  plates at very early universe.
\end{abstract}
\newpage
% \vspace*{10mm}

\section{Introduction}
The Casimir effect is one of the most interesting manifestations
  of non-trivial properties of the vacuum state in quantum field
  theory [1,2]. Since its first prediction by
  Casimir in 1948\cite{Casimir} this effect has been investigated for
  different fields having different boundary geometries \cite{{Remeo},{setare0},{setare},{setare1},{setare2},{setare3},
  {setare4},{SET}}. The
  Casimir effect can be viewed as the polarization of
  vacuum by boundary conditions or geometry. Therefore, vacuum
  polarization induced by a gravitational field is also considered as
  a Casimir effect.\\
  Casimir effect may  have interesting implications for the early
  universe. In \cite{set3} the
   Casimir effect of a massless scalar field with Dirichlet
   boundary condition in a spherical shell having different vacua
   inside and outside, which represents a bubble in the early universe with
   false/true vacuum inside/outside, has been investigated. The Casimir stress on two
    concentric spherical
   shells with constant comoving radius having different vacua inside and
    outside in de Sitter space, which corresponds to a
    spherically symmetric domain wall with thickness, has been calculated in \cite{set4}.\\
  In the context of hot big bang cosmology, the unified theories
  of the fundamental interactions predict that the universe passes
  through a sequence of phase transitions. Different types of topological objects may
  have been formed during these phase transitions, these include domain walls, cosmic
strings and monopoles \cite{{zel},{kib},{viel}}. These topological
defects appear as a consequence of breakdown of local or global
gauge symmetries of a system composed by self-coupling iso-scalar
Higgs fields
$\Phi^a$.\\
On the other hand, signature changing space-times have recently been
of particular importance as the specific geometries with interesting
physical effects. The original idea of signature change was due to
Hartle, Hawking and Sakharov \cite{{HHS},{HHS1}}. This interesting
idea would make it possible to have  both Euclidean and Lorentzian
metrics in path integral approach to quantum gravity. Latter it was
shown that the signature change may happen even in classical general
relativity, as well \cite{CSC}-\cite{CSC8}. The issue of propagation
of quantum fields on signature-changing space-times has also been of
some interest \cite{D}-\cite{D5}. For example, Dray {\it et al} have
shown that the phenomenon of particle production may happen for
scalar particles propagation in space-time with heterotic signature.
They have also obtained a rule for propagation of massless scalar
fields on a two- dimensional space-time with signature change.
Dynamical determination of the metric signature in space-time of
nontrivial topology is another interesting issue which has been
studied in Ref.\cite{eor}.\\
To the author knowledge, no attempt has been done to study the
Casimir effect within the geometries with signature change. A
relevant work to the present paper is Ref. \cite{FM}. In this work
the Casimir effect for the free massless scalar field propagating on
a two-dimensional cylinder with a signature-changing strip has been
studied. Motivated by this new element in studying the Casimir
effect we have paid attention to study such a non-trivial effect in
a model of two parallel plates in de Sitter space with different
cosmological constants and metric signatures between and outside.
\section{Parallel Plates with Robin boundary conditions in de Sitter Space }

Consider a massless scalar field coupled conformally to a de Sitter
background space. The scalar field satisfies the following Robin
boundary conditions on two parallel plates within an arbitrary
space-time \cite{sah}:
 \begin{equation}\label{boun}
 (1+\beta_{m}(-1)^{m-1}
 \partial_{x})\phi(x^{\nu})|_{x = a_{m}} = 0, \hspace{2cm}  m = 1,2,
  \end{equation}
Here we have assumed that the two plates are normal to the
Cartesian $x$-axis at $x = a_{1, 2}$.\\
 The Robin boundary condition may be interpreted as the boundary condition on a thick
plate \cite{leb}. Rewriting(1) in the following form:
\begin{equation}
\partial_{x}\phi(x^{\nu})=(-1)^{m}\frac{1}{\beta_{m}}\phi(x^{\nu}),
\end{equation}
where $|\beta_{m}|$, having the dimension of a length, may be called
the skin-depth parameter. This is similar to the case of penetration
of an electromagnetic field into a real metal, where the tangential
component of the electric
field is proportional to the skin-depth parameter.\\
 The corresponding
field equation has the form
\begin{equation}
(\nabla_{\mu}\nabla^{\mu}+\xi R)\phi=0, \hspace{1cm}
\nabla_{\mu}\nabla^{\mu}=\frac{1}{\sqrt{-g}}\partial_{\mu}(\sqrt{-g}g^{\mu\nu}\partial_{\nu}),
\label{eqmo}
\end{equation}
where $\nabla_{\mu}$ is the covariant derivative operator, and $R$
is the Ricci scalar for de Sitter space. In the conformally
coupled case, the corresponding stress-energy tensor is defined
as \cite{davies}
\begin{equation}
T_{\mu\nu}=(1-2\xi)\partial_{\mu}\phi\partial_{\nu}\phi+(2\xi-1/2)g_{\mu\nu}\partial^{\lambda}\phi\partial_{\lambda}
\phi-2\xi\phi\nabla_{\mu}\nabla_{\nu}\phi+\frac{1}{12}g_{\mu
\nu}\phi\nabla_{\lambda}\nabla^{\lambda}\phi, \label{enmom}
\end{equation}
where $\xi=\frac{1}{6}$. It is known that in the Minkowski
space-time for the conformally coupled scalar field the
perpendicular pressure, $P$, is uniform in the region between the
plates and is given by \cite{sah}
\begin{equation}
 P=3\varepsilon_{c},
\end{equation}
where $\varepsilon_{c}$ is the Casimir energy density. This
Casimir energy has been calculated to be
\begin{equation}
\varepsilon = \varepsilon_{c}=
\frac{-A}{8\Gamma(5/2){\pi}^{3/2}a^{4}},
\end{equation}
where $A$ depends on $\beta_{1, 2}$ and $a$ may be inferred from
Eq.(4.15) of \cite{sah}. It has also been shown that for $\beta_{1}
= -\beta_{2}$
\begin{equation}
\varepsilon=\varepsilon_{c}=
\frac{-\zeta_{R}(4)\Gamma(2)}{(4\pi)^{2}a^{4}}=\frac{-\pi^{2}}{1440a^{4}},
\end{equation}
which is the same as for the Dirichlet and Neumann boundary
conditions.\\
 Consider now two parallel plates in a de Sitter
space-time. To make the maximum use of the flat-space calculations
we use the de Sitter metric in the conformally flat form:
\begin{equation}\label{6}
ds^{2}=\frac{\alpha^{2}}{\eta^{2}}
[d\eta^{2}-\sum_{i=1}^{3}(dx^{i})^{2}],
\end{equation}
where $\eta$ is the conformal time:
\begin{equation}
-\infty < \eta < 0.
\end{equation}
The relation between the parameter $\alpha$ and the cosmological
constant $\Lambda$ is given by
\begin{equation}
\alpha^{2}=\frac{3}{\Lambda}.
\end{equation}
The quantization of a scalar field on background of the metric
  (\ref{6}) is straightforward. Let $\{\phi_{k}(x), \phi^{\star}_{k}(x)\}$
  be a complete set of orthonormalized positive and negative
  frequency solutions to the field equation (\ref{eqmo}), obeying
  boundary condition (\ref{boun}). By expanding the field operator
  over these eigenfunctions, using the standard commutation rules
  and the definition of the vacuum state for the vacuum
  expectation values of the energy-momentum tensor one obtains
  \cite{davies}
\begin{equation}
\langle0|T_{\mu}^{\nu}|0\rangle=\sum_{k}T_{\mu}^{\nu}\{\phi_{k},
\phi_{k}^{*}\}, \label{19}
\end{equation}
where $|0>$ is the amplitude for the corresponding vacuum state,
and the bilinear form $T_{\mu}^{\nu}\{\phi,\phi^{*} \}$ on the
right is determined by the classical energy-momentum tensor
(\ref{enmom}). In the problem under consideration we have a
conformally trivial situation, namely a conformally invariant
field on background of the conformally flat spacetime. Instead of
evaluating Eq.(\ref{19}) directly on background of the curved
metric, the vacuum expectation values can be obtained from the
corresponding flat spacetime results for a scalar field $\phi$ by
using the conformal properties of the problem under consideration.
Under the conformal transformation $\tilde
g_{\mu\nu}=\Omega^{2}g_{\mu\nu}$ the $\phi$ field will be changed
by the rule
\begin{equation}\label{fieltra}
\tilde\phi(x)=\Omega^{-1}{\phi}(x).\label{*}
\end{equation}
Using the standard relation between the energy-momentum tensor
for conformally coupled situations \cite{davies}
\begin{equation}\label{15}
<T^{\mu}_{\nu}[\tilde
g_{\alpha\beta}]>=(\frac{g}{\tilde{g}})^{\frac{1}{2}}
<T^{\mu}_{\nu}[{g^{(M)}_{\alpha\beta}}]>-\frac{1}{2880\pi^{2}}[\frac{1}{6}
\tilde H^{(1)\mu}_{\nu}-\tilde H^{(3)\mu}_{\nu}],
\end{equation}
where $g_{\mu \nu}$ and $\tilde g _{\mu \nu}$ are conformal to each
other, with their respective determinants $g$ and $\tilde{g}$. We
are going to assume that $g_{\mu \nu}$ is the Minkowski metric. Now,
$<T^{\mu}_{\nu}[{g^{(M)}_{\alpha\beta}}]>$, the regulatized energy-
momentum tensor for a conformally coupled scalar field in the case
of a parallel plate configuration in flat space-time is given by
\begin{equation}\label{13}
<T^{\mu}_{\nu}[{g^{(M)}_{\alpha\beta}}]>=diag(\varepsilon,-P,-P_{\bot},-P_{\bot})
=diag(\varepsilon,3\varepsilon,-\varepsilon,-\varepsilon).
\end{equation}
The second term in (\ref{15}) is the vacuum polarization due to the
gravitational field, without any boundary conditions. The functions
$H^{(1,3)\mu}_{\nu}$ are some combinations of curvature tensor
components (see \cite{davies}). For a massless scalar field in de
Sitter space, the term is given by \cite{{davies},{Dowk}}
\begin{equation}\label{11}
-\frac{1}{2880}[\frac{1}{6} \tilde H^{(1)\mu}_{\nu}-\tilde
H^{(3)\mu}_{\nu}]
=\frac{1}{960\pi^{2}\alpha^{4}}\delta^{\mu}_{\nu}.
\end{equation}
From(5,7,10,13,14) one can obtain the vacuum pressure due to the
boundary acting on the plates:
\begin{equation}\label{12}
P^{(1,2)}_{B}=P_{B}(x_{1,2})=(\frac{g}{\tilde{g}})^{\frac{1}{2}}
(\frac{-3\pi^{2}}{1440 a^{4}})=\frac{\eta^{4}}{\alpha^{4}}
(\frac{-3\pi^{2}}{1440 a^{4}})=\frac{-\eta^{4}\Lambda^{2}}{3}
\frac{\pi^{2}}{1440 a^{4}},
\end{equation}
which is attractive. It has been shown that this pressure is zero
for $x < a_1$ and $x > a_2$ \cite{{set1}, {set2}}. The
gravitational part of the pressure according to (\ref{11}) is
equal to
\begin{equation}
P_{G}=-<T^{1}_{1}>=\frac{-1}{960\pi^{2}\alpha^{4}}.
\end{equation}
This is the same from both sides of the plates, and hence leads to
zero effective force. Therefore, the effective force acting on the
plates is only given by the boundary part.

\section{Parallel plates with different cosmological constants and signatures
 between and outside}
Now, assume that there are different vacua between and outside of
 the plates, corresponding to $\alpha_{betw}$ and $\alpha_{out}$ in the
 metric(8). As we have seen in the previous section, the vacuum
 pressure due to the boundary is only non-vanishing between the
 plates. Therefore, we have for the pressure due to the boundary
 \begin{equation}
 P^{(1,2)}_{B}=\frac{\eta^{4}}{\alpha_{betw}^{4}}\frac{-3\pi^{2}}{1440a^{4}}
 =\frac{-\eta^{4}\Lambda_{betw}^{2}}{3}\frac{\pi^{2}}{1440a^{4}}.
 \end{equation}
Now, we obtain the pure effect of vacuum polarization due to the
gravitational field without any boundary conditions in Euclidean
(outside) region with the following metric
\begin{equation}
ds^{2}=-\frac{\alpha^{2}}{\eta^{2}}[d\eta^{2}+\sum_{i=1}^{3}(dx^{i})^{2}].
\label{14}
\end{equation}
To this end, we calculate the renormalized stress tensor for the
massless scalar field in de Sitter space with Euclidean signature.
We use Eq.(\ref{15}), then we obtain
$<T^{\mu}_{\nu}[{g^{(M)}_{\alpha\beta}}]>=0$, and
$$
^{(1)}H^{\mu}_{\nu}=0,
$$
$$
^{(3)}H^{\mu}_{\nu}=\frac{3}{\alpha^4}\delta^{\mu}_{\nu}.
$$
Therefore
\begin{equation}
<T^{\mu}_{\nu}[\tilde g_{\alpha\beta}]>=\frac{1}{960 \pi^2
\alpha^4}\delta^{\mu}_{\nu}, \label{16}
\end{equation}
which is exactly the same result for the Lorentzian case
\cite{davies}. The corresponding effective pressures for the
Euclidean (outside) and Lorentzian (between) regions with
$\alpha_{out}$ and $\alpha_{betw}$, due to pure effect of
gravitational vacuum polarization without any boundary condition,
are given respectively by
\begin{equation}\label{out}
 P^{E}_{out}=-<T^{1}_{1}>_{out}=\frac{-1}{960\pi^{2}\alpha_{out}^{4}}=
 \frac{-\Lambda_{out}^{2}}{9}\frac{1}{960\pi^{2}}.
 \end{equation}
\begin{equation}\label{betw}
 P^{L}_{betw}=-<T^{1}_{1}>_{betw}=\frac{-1}{960\pi^{2}\alpha_{betw}^{4}}=
 \frac{-\Lambda_{betw}^{2}}{9}\frac{1}{960\pi^{2}},
 \end{equation}
The corresponding gravitational pressure on the plates is then
given by
\begin{equation}
P_G = P_{betw}^{L}-P_{out}^{E} = \frac{-1}{9\times 960\pi^{2}}
 (\Lambda_{betw}^{2}-\Lambda_{out}^{2}).
\label{17}
\end{equation}
We now proceed to calculate the stress due to the boundary effects
$P_B$. The stress on the plates due to boundary effects for the
Lorentzian metric (\ref{6}) has been obtained as (\ref{12}). In
signature changing case we have correspondingly
\begin{equation}
P_{B}^{L-E}=\langle0|T_{x}^{x}|0\rangle|^{L}_{betw}-\langle0|T_{x}^{x}|0\rangle|^{E}_{out}.
\label{18}
\end{equation}
The scalar field $\phi(\vec{x}, \eta)$ in the Lorentzian de Sitter
space satisfies
\begin{equation}
(\Box+\xi R)\phi(\vec{x}, \eta)=0, \label{21}
\end{equation}
where $\Box$ is the Laplace-Beltrami operator for de Sitter
metric, and $\xi$ is the coupling constant. For conformally
coupled field in four dimension $\xi=\frac{1}{6}$, and R , the
Ricci scalar curvature, is given by
\begin{equation}
R=12\alpha^{-2}.
\end{equation}
Taking into account the separation of variables as
\begin{equation}
\phi(\vec{x},\eta)=A(\vec{x})T(\eta), \label{22}
\end{equation}
for the inside Lorentzian domain with
\begin{equation}
T_L(\eta)=\frac{1}{\sqrt{2\omega(2\pi)^3}}\exp^{-i\omega\eta},
\label{23}
\end{equation}
the corresponding Euclidean $\eta$-dependence takes on the form
\begin{equation}
T_E(\eta)=\frac{1}{\sqrt{2\omega(2\pi)^3}}\exp^{-\omega\eta},
\label{24}
\end{equation}
for the scalar field to be normalizable in $\eta$. Based on the
following mode expansion
\begin{equation}
\phi=\sum_{i}(a_i^-\varphi_i^-+a_i^+\varphi_i^+),
\end{equation}
and normal ordering $\langle0|a_i^- a_i^+ |0\rangle=1$,  the
detailed calculations, using Eqs.(\ref{19}),(\ref{23}) and
(\ref{24}) in Eq.(\ref{18}), lead to
\begin{equation}\label{25}
P_{B}^{L-E}=\frac{-\eta^{4}\Lambda_{betw}^{2}}{3}\frac{\pi^{2}}{1440a^{4}}+\frac{\eta^2}{576\pi^5}(\Lambda_{betw}^2-
\Lambda_{out}^2)\zeta(2).
\end{equation}
where $\zeta(2)$ is the Zeta function and Abel-Plana summation
formula has been used to regularize the infinite sum
$\sum_{i}\omega_i$. Taking into account the gravitational pressure
on the plates we obtain the total result
\begin{equation}
P=P_G+P_B= \frac{-1}{9\times 960\pi^{2}}
 (\Lambda_{betw}^{2}-\Lambda_{out}^{2})-\frac{\eta^{4}\Lambda_{betw}^{2}}{3}\frac{\pi^{2}}{1440a^{4}}
 +\frac{\eta^2}{576\pi^5}(\Lambda_{betw}^2-
\Lambda_{out}^2)\zeta(2) \label{26}
\end{equation}
The first term $P_G$ which is pure gravitational effect without
boundary is $\eta$-independent, but is sensitive to the initial
(in terms of $\eta$) values of $\Lambda_{betw}$ and
$\Lambda_{out}$. Given a false
 vacuum between the plates, and true vacuum out-side, i.e. $\Lambda_{betw}>
 \Lambda_{out}$, then the gravitational part is negative, the second term is always negative, but in
 this case the last term is positive. At very early universe, namely $\eta \simeq 0$ the
constant first term dominates and there is an attraction between
the plates.  On the other hand, at times $\eta >1$, the first
term is ignorable and there is a competition between the second
(negative) and third (positive) terms. For the case of true
vacuum between
 the plates and false vacuum out-side, i.e.
 $\Lambda_{betw}<\Lambda_{out}$, the gravitational pressure is
 positive, and another terms are negative. Therefore, the total pressure may be either negative or
 positive. In this case, at very early universe, i.e. $\eta \simeq
 0$, the total pressure  $P>0$, this initial repulsion of the parallel plates
 will stopped at times $\eta >1$.
 \section{Conclusion}
In the present paper we have investigated the Casimir effect for a
conformally coupled massless scalar field confined in the region
between two parallel plates in a de Sitter background corresponding
to different metric signatures and cosmological constants between
and outside of the plates.  The general case of the mixed(Robin)
boundary conditions is considered. The vacuum expectation values of
the energy-momentum tensor are derived from the corresponding flat
spacetime results by using the conformal properties of the problem.
Previously this method has been used in \cite{SET} to investigate
the vacuum characteristics of the Casimir configuration on a
background of conformally flat brane-world geometries for a massless
scalar field with Robin boundary conditions on plates. Also this
method has been used in \cite{set1} to derive the vacuum
characteristics of the Casimir configuration on a background of
static domain wall geometry for a scalar field with Dirichlet
boundary condition on plates. (For investigations of the Casimir
energy in brane-world models with de Sitter branes, see
    Refs. \cite{fab}-\cite{fab5}).\\
Our calculation
 shows that for the parallel plates with false vacuum between and true
  vacuum outside, the total Casimir pressure leads to an attraction of the
  plates at very early universe, namely $\eta \simeq 0$. The boundary term is proportional to
   the fourth power of the inverse
 distance between the plates, and is always negative, which means a huge
 attractive force for small distances. In contrast, plates with a true
 vacuum between them may repel each other to a maximum distance
 and attract again. The result may be of interest in the case of
 formation of cosmic domain walls in the early universe, where the
  wall orthogonal to the $x-$axis is described by the
 function $\phi_{i}(x)$ interpolating between two different minima
 at $x\rightarrow \pm\infty$ \cite{vil1}. Also our calculations may be of interest in the brane-world
cosmological scenarios. The brane-world corresponds to a manifold
with boundaries and all fields which propagate in the bulk will give
Casimir-type contributions to the vacuum energy, and as a result to
the vacuum forces acting on the branes. In dependence of the type of
a field and boundary conditions imposed, these forces can either
stabilize or destabilize the brane-world. In addition, the Casimir
energy gives a contribution to both the brane and bulk cosmological
constant and, hence, has to be taken into account in the
self-consistent formulation of the brane-world dynamics. In general,
we believe the idea of Casimir effect in signature-changing
space-time is novel and interesting. Therefore, the study of Casimir
effect in the present model may provide important results relevant
to the study of cosmological  brane-world scenarios.
  
\end{document}